\documentclass[twocolumn,amsmath,showpacs,amsfonts,aps,prc,floatfix]{revtex4}
\usepackage{graphicx}
\usepackage{bm}

\begin{document}


\title{Multiplicity, mean $p_T$, $p_T$-spectra and elliptic flow of identified particles in Pb+Pb collisions at LHC}
\author{A. K. Chaudhuri}
\email[E-mail:]{akc@veccal.ernet.in}
\affiliation{Variable Energy Cyclotron Centre, 1/AF, Bidhan Nagar, 
Kolkata 700~064, India}

\begin{abstract}

Israel-Stewart's causal theory of dissipative hydrodynamics, with the
ADS/CFT lower limit of shear viscosity to entropy ratio ($\eta/s$=0.08),
give consistent description of a number of experimental observables in Au+Au collisions at RHIC (c.m. energy $\sqrt{s}$=200 GeV) \cite{Chaudhuri:2008sj}.  Assuming that in
Pb+Pb collisions at LHC (c.m. energy $\sqrt{s}$=5.5 TeV), except for the initial temperature,
other parameters of the fluid remain unchanged,  we have predicted for the centrality dependence of  multiplicity, mean $p_T$, $p_T$-spectra, elliptic flow.   
The central temperature of the fluid   is  adjusted to $T_i$=421 MeV such that  in a Pb+Pb collision, with participant number $N_{part}$=350, average charge particle multiplicity is    $\sim$ $900$ and is consistent with the experimental trend observed at lower energies. Compare to Au+Au collisions at RHIC, in Pb+Pb collisions at LHC,   
on the average, particle multiplicity increases by a factor of $\sim$1.6  , the mean $p_T$ is increased by $\sim$10\% only. The elliptic flow on the other hand decreases by $\sim$15\%.
\end{abstract}


\date{\today}  

\maketitle

\section{Introduction} \label{sec1}
Experiments in Au+Au collisions at RHIC  \cite{BRAHMSwhitepaper,PHOBOSwhitepaper,PHENIXwhitepaper,STARwhitepaper}, produced convincing evidences that in non-central Au+Au collisions, a hot, dense, strongly interacting,  collective QCD matter is created. Whether the matter can be characterized as the lattice QCD \cite{lattice} predicted Quark-Gluon-Plasma (QGP) or not,   is still a question of debate. 
 Relativistic hydrodynamics provides a convenient tool to analyse Au+Au collision data. It is assumed that in the collision a fireball is produced. Constituents of the fireball collide frequently to establish local thermal equilibrium sufficiently fast and after a certain time $\tau_i$, hydrodynamics become applicable. If the macroscopic properties of the fluid e.g. energy density, pressure, velocity etc. are known at the equilibration time $\tau_i$, the relativistic hydrodynamic equations can be solved to give the space-time evolution of the fireball till a given freeze-out condition such that interactions between the constituents are too weak to continue the evolution. 
Using suitable algorithm (e.g. Cooper-Frye) information at the freeze-out can be converted into particle spectra and can be directly compared with experimental data. Thus, hydrodynamics, in an indirect way, can characterize the initial condition of the medium produced in heavy ion collisions. Hydrodynamics equations are closed only with an equation of state  and one can investigate the possibility of phase transition in the medium. 
 
A host of experimental data produced in Au+Au collisions at RHIC, at c.m. energy $\sqrt{s}$=200 GeV, have been successfully analysed using ideal hydrodynamics \cite{QGP3}. Multiplicity, mean $p_T$, $p_T$-spectra, elliptic flow etc.  of identified particles, are well explained in the ideal hydrodynamic model with QGP as the initial state.  Ideal hydrodynamics analysis of the RHIC data indicate that in central Au+Au collisions, at the equilibration time $\tau_i \approx$ 0.6 fm, 
  central energy density of the QGP fluid is $\varepsilon_i \approx$30 $GeV/fm^{-3}$ \cite{QGP3}. It may be mentioned that ideal hydrodynamics  description of data are not unblemished. 
  $p_T$ spectra or the elliptic flow are explained only up to  transverse momenta $p_T \approx 1.5 GeV$. At higher $p_T$ description deteriorates. Also ideal hydrodynamic description to data gets poorer in peripheral collisions.

Hydrodynamics implicitly assume local thermal equilibration. In the rest frame of the fluid, particle momenta are isotropic. But
at the early stage of the evolution, assumption of isotropic momentum distribution cannot be very accurate.  
System could only be in partial  equilibration and for a creditable analysis of the experimental data, dissipative effects  must be accounted for. 
Shear viscosity is the most important dissipative effect in heavy ion collisions. Shear viscosity of QGP matter is quite uncertain. In recent years, string theory motivated calculations \cite{Policastro:2001yc} indicated that shear viscosity over entropy
ratio of any matter is bounded from the lower side, $\eta/s \geq 1/4\pi$. It is then expected that at the minimum, shear viscosity over entropy of QGP matter should be $1/4\pi$.  

In recent years there has been significant progress in numerical implementation of viscous dynamics  \cite{Teaney:2003kp,MR04,Koide:2007kw,Chaudhuri:2005ea,Heinz:2005bw,asis,Chaudhuri:2008sj,Romatschke:2007mq,Romatschke:2007jx,Baier:2006gy,Song:2007fn,Song:2007ux} . At the Cyclotron Centre, Kolkata, we have developed a code "AZHYDRO-KOLKATA", to solve Israel-Stewart's 2nd order theory for dissipative hydrodynamics in 2+1 dimensions.  
In a recent publication \cite{Chaudhuri:2008sj}, it was shown that minimally viscous hydrodynamics ($\eta/s$=0.08), with QGP as the initial state consistently explain a large part of RHIC data, e.g. $p_T$-spectra of identified particles, elliptic flow in minimum-bias/mid-central collisions etc. 
Indeed, the description is better than that obtained in ideal dynamics, particularly at large $p_T$.  
 
In not too distant future, Large Hadron Collider (LHC) is expected to be operational. It is being planned to collide Lead beams at c.m. energy $\sqrt{s}$=5.5 TeV, which is 27 times larger than the RHIC energy collisions. One expects that  conclusive evidence for QGP formation can be obtained in Pb+Pb collisions at LHC. 
In the present paper, in the framework of minimally viscous hydrodynamics, we have given predictions for several experimental observables, which will be measured in Pb+Pb collisions at LHC energy.  
It will be shown that even if, from RHIC to LHC collisions, energy is increased by a factor of  27,  the experimental observables e.g. multiplicity, $p_T$ spectra, mean $p_T$, elliptic flow etc. do show rather modest variation from RHIC energy collisions. The reason can be understood also.  Multiplicity or entropy of the system increases logarithmically with energy. Thus, from RHIC to LHC,  even though collision energy increases by a factor 27, the entropy or multiplicity increase by a factor of $\sim$ 1.6 only.  Experimental observables do not show very large variation from RHIC energy collisions. 

\begin{table}[h]
\begin{ruledtabular} 
  \begin{tabular}{|c|cc|cc|}\hline
    &    PB+Pb@LHC    & & Au+Au@RHIC &  \\
centrality & $<N_{part}>$ & $<b>$ & $<N_{part}>$ & $<b>$ \\  \hline
   0-10       & 346.8     &   3.51 &318.9&3.30   \\ \hline
   10-20       &245.1      &  6.13 &223.1& 5.79   \\ \hline
    20-30      & 169.9     &  7.93  &154.1 &7.49  \\ \hline
     30-40     &  113.4    &  9.38  &102.8 & 8.87  \\ \hline
     40-50     &  71.6    &    10.64 & 65.0 &10.06 \\ \hline
     50-60     &  41.5    &    11.76  &  37.8 &11.11\\ \hline
     60-70     &  21.5    &    12.79  & 19.8 & 12.09 \\ \hline
      70-80    &   9.6   &   13.76  &9.08 & 13.0  \\ \hline
       80-90   &    3.1  &    14.80  & 2.97 & 14.03  
  \end{tabular}
\end{ruledtabular}  
\caption{\label{table1} Glauber model calculation for the average participant number 
$<N_{part}>$ and average impact parameter $<b>$ (in fm) for different ranges of centrality cuts in Pb+Pb/Au+Au  collisions at LHC/RHIC energy.} 
\end{table} 
 
\section{initial conditions for Pb+Pb collisions at LHC}

Details of our solution of 
Israel-Stewart's \cite{IS79} 2nd order theory of dissipative hydrodynamics 
  can be found in  \cite{Chaudhuri:2008sj}. Briefly,
assuming longitudinal boost-invariance, we have solved the Israel-Stewart's 2nd order theory for a baryon free fluid with dissipation due to shear viscosity only. 
In Israel-Stewart's theory, dissipative flows are treated as  extended thermodynamic variables.  For a baryon free fluid with only shear viscosity as the dissipative effect, energy-momentum conservation equation is required to be solved simultaneously with the relaxation equation for the shear stress tensor,  

\begin{eqnarray}  
\partial_\mu T^{\mu\nu} & = & 0, \label{eq1} \\
D\pi^{\mu\nu} & = & -\frac{1}{\tau_\pi} (\pi^{\mu\nu}-2\eta \nabla^{<\mu} u^{\nu>}). \label{eq2}
\end{eqnarray}

 Eq.\ref{eq1} is the conservation equation for the energy-momentum tensor, $T^{\mu\nu}=(\varepsilon+p)u^\mu u^\nu - pg^{\mu\nu}+\pi^{\mu\nu}$, 
$\varepsilon$, $p$ and $u$ being the energy density, pressure and fluid velocity respectively. Eq.\ref{eq2} is the relaxation equation for the shear stress tensor $\pi^{\mu\nu}$.   
In Eq.\ref{eq2}, $D=u^\mu \partial_\mu$ is the convective time derivative, $\nabla^{<\mu} u^{\nu>}= \frac{1}{2}(\nabla^\mu u^\nu + \nabla^\nu u^\mu)-\frac{1}{3}  
(\partial . u) (g^{\mu\nu}-u^\mu u^\nu)$ is a symmetric traceless tensor. $\eta$ is the shear viscosity and $\tau_\pi$ is the relaxation time. In \cite{Chaudhuri:2008sj}
Eqs.\ref{eq1} and \ref{eq2}  are solved in $(\tau=\sqrt{t^2-z^2},x,y,\eta_s=\frac{1}{2}\ln\frac{t+z}{t-z})$ coordinates, assuming boost-invariance.  

We note that presently there is disagreement about the form of the relaxation equation to be used in heavy ion collisions.   
   In \cite{Romatschke:2007mq,Song:2007fn} an extra term $R=[u^\mu\pi^{\nu\lambda}+u^\nu\pi^{\nu\lambda}]Du_\lambda$ is included in the relaxation equation. Shear stress tensor is traceless ($\pi^\mu_\mu=0$) and transverse to 4-velocity ($u_\mu \pi^{\mu\nu}=0$). The term is needed to maintain the transversality  and tracelessness  condition. Israel-Stewart \cite{IS79} developed the theory on gradient expansion of entropy,
gradients of equilibrium thermodynamical variables are assumed to be small.
The  term $[u^\mu\pi^{\nu\lambda}+u^\nu\pi^{\nu\lambda}]Du_\lambda$ does not contribute to entropy and is missed
in Israel-Stewart's theory. Moreover, in Israel-Stewart's theory, the term can be neglected (both $\pi^{\mu\nu}$ and $Du_\mu$ are small and their combination is neglected). We have also checked that for minimally viscous fluid, contribution of the
term is negligible. For minimally viscous fluid, energy density evolution is hardly affected whether the term is present or not in the relaxation equation  (see Fig.8 and 9 of \cite{Chaudhuri:2008sj}).

Details of the analysis of RHIC data in Au+Au collisions can be found in \cite{Chaudhuri:2008sj}. In brief, minimally viscous  QGP fluid was initialised with a Glauber model initial condition, with 75\% soft collisions and 25\% hard collisions. In a b=0 collision, this corresponds to 
central entropy density $S_{ini}$=110 $fm^{-3}$, at the initial time $\tau_i$=0.6 fm. The transverse fluid velocity at the initial time was assumed to be zero, $v_x=v_y=0$. The shear stress tensor $\pi^{\mu\nu}$ was assumed to attain boost-invariant value at the initial time $\tau_i$. 
For the relaxation time $\tau_\pi$, Boltzmann approximation $\tau_\pi=3\eta/2p$ is used.
The freeze-out temperature was varied to fit elliptic flow in 16-23\% centrality Au+Au collisions. It was seen that for $T_F$=130 MeV, elliptic flow in 16-23\% centrality Au+Au collisions as well as a host of other data are explained. For the equation of state we have used EOS-Q developed in \cite{QGP3}, with bag model EOS for the QGP phase and hadronic resonance gas for the hadronic phase. EOS-Q has a 1st order phase transition at $T_c$=164 MeV. 
 
We assume
that in Pb+Pb collisions at LHC, except for the central entropy density, other parameters of the model remain unchanged. In Pb+Pb collisions also, the QGP fluid is thermalised at the same time as in Au+Au collisions at RHIC i.e. $\tau_i$=0.6 fm. The initial fluid velocity is zero: $v_x=v_y=0$, and the shear stress tensor has attained the boost-invariant value. And as in Au+Au collisions, in Pb+Pb collisions also, the hadronic fluid freezes-out at $T_F$=130 MeV.  
The central energy density or entropy density in Pb+Pb 
collisions at LHC energy ($\sqrt{s}$=5.5 TeV) cannot be same as in Au+Au collisions at RHIC ($\sqrt{s}$=200 GeV).
One expects larger energy deposition in Pb+Pb collisions.
To obtain the initial energy/entropy density of the fireball in LHC energy collisions we proceed as follows:

\begin{figure}[t]
\includegraphics[bb=55 291 544 770
,width=0.5\linewidth,clip]{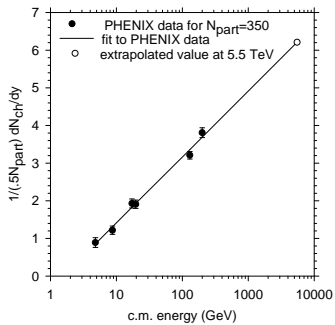}
\caption{Filled circles are the PHENIX data for the charged particle multiplicity per participant $\frac{1}{.5N_{part}}\frac{dN}{dy}$ as a function of c.m. energy for participant number $N_{part}$=350. The solid line is a fit to the 
PHENIX data by Eq.\ref{eq3}. The unfilled circle is the extrapolated value of $\frac{1}{.5N_{part}}\frac{dN}{dy}$    at LHC energy $\sqrt{s}$=5.5 TeV, for participant number $N_{part}=350$.}
\label{F1}
\end{figure} 

PHENIX collaboration \cite{Adler:2004zn} has tabulated the average charged particle multiplicity as
a function of collision energy for a range of collision centrality. 
In Fig.\ref{F1}, for participant number $N_{part}$=350, the average multiplicity $\frac{1}{.5N_{part}} \frac{dN_{ch}}{d\eta}$ is shown as a function of collision energy . The multiplicity increases logarithmically with energy,

\begin{equation} \label{eq3}
\frac{dN_{ch}}{d\eta}=A+ B \ln \sqrt{s},
\end{equation}

\noindent with $A=-0.33$ and $B=0.75$. We use the relation to extrapolate to LHC energy $\sqrt{s}$=5.5 TeV. The extrapolated value of
average charged particle multiplicity in LHC energy is $\sim 927\pm 70$.   
 We adjust
the central entropy density to $S_{ini}$=180 $fm^{-3}$ such that a $N_{part}$=350 Pb+Pb collision produce $\sim$ 900 charged particles. Entropy density $S_{ini}$=180 $fm^{-3}$ corresponds to central temperature   $T_i$=421 MeV.
Compared to Au+Au collisions at RHIC (central temperature $T_i$=357 MeV),  in Pb+Pb collisions at LHC, central temperature is $\sim$ 20\% higher.  
 
With the initial condition as described above we have solved the hydrodynamic
equations and calculate invariant particle yield from the freeze-out surface at $T_F$=130 MeV. It may be noted that we are assuming boost-invariance. Consequently, our predictions are valid only in the mid-rapidity range.
In the following, we will show our predictions as a function of collision centrality or rather as a function of number of participants.
Glauber model calculations for average participant number $<N_{part}>$ and average impact parameter $<b>$ for different centrality cuts are given in table \ref{table1}. For Pb+Pb collisions at LHC energy, we have used $\sigma_{inel}$=70 mb. For comparison, in table \ref{table1}, we have also shown the same results for Au+Au collisions at RHIC, when $\sigma_{inel}$=44 mb. 

\begin{figure}[t]
\includegraphics[bb=28 287 524 769
,width=0.5\linewidth,clip]{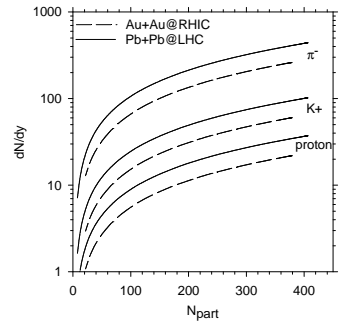}
\caption{ 
The black, red and green solid and dashed lines are minimally viscous hydrodynamics predictions for the centrality dependence of $\pi^-$, $K^+$ and proton multiplicity in Pb+Pb collisions at LHC and in Au+Au collisions at RHIC.
 The yileds are normalised by a factor of N=1.4 to account for the neglect of resonance production.} 
\label{F2}
\end{figure}

\section{multiplicity and mean $p_T$ in Pb+Pb collisions at LHC}

Particle multiplicity is one of the important observables in heavy ion collisions. It is a measure of the entropy of the system. 
In Fig.\ref{F2}, 
the black, red and green lines are minimally viscous hydrodynamic
model predictions for the centrality dependence of   $\pi^-$, $K^+$ and proton multiplicity in Pb+Pb collisions at LHC. For comparison, minimally viscous hydrodynamic
predictions for $\pi^-$, $K^+$ and proton multiplicity in   Au+Au collisions at RHIC are also shown (the dashed lines) in Fig.\ref{F2}.
We have neglected resonance production. To account for the neglect of resonance
production, yields are normalised by a factor of $N=1.4$. 
From the predictions, it appears that compared to Au+Au collisions,
in central/mid-central Pb+Pb collisions, particle yields are enhanced by a factor of $\sim$ 1.6-1.8 . It is expected. Multiplicity increases logarithmically with energy.   
 
 \begin{figure}[t]
\includegraphics[bb=47 312 538 793
,width=0.5\linewidth,clip]{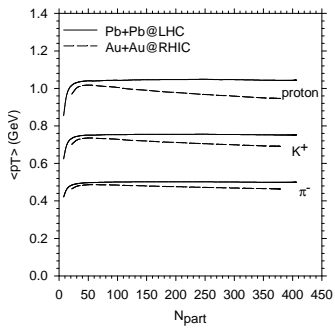}
\caption{    The   solid lines are predicted mean $p_T$ for $\pi^-$, $K^+$ and proton in Pb+Pb collisions at LHC energy. For comparison, minimally viscous hydrodynamics predictions for mean $p_T$ in Au+Au collisions are also shown by the dashed lines.}
\label{F3}
\end{figure}
 
 Centrality dependence of mean $p_T$ is another important observable. 
 In Au+Au collisions at RHIC, in minimally viscous hydrodyanmics,
 centrality dependence of $<p_T>$ of identified particles are reasonably well expalined. 
In Fig.\ref{F3} minimally hydrodynamics predictions
 for mean $p_T$ for $\pi^-$, $K^+$ and protons, in Pb+Pb colliisons at LHC are shown. In collisions beyond $N_{part}$=50, mean $<p_T>$   is approximately constant; $<p_T> \approx$ 0.5, 0.75 and 1 for $\pi^-$, $K^+$ and protons respectively. For comparison, predictions for mean $p_T$ in Au+Au collisions at RHIC are  shown in Fig.\ref{F3} as the dashed lines. 
From RHIC to LHC, even though collision energy is increased by a factor of 27, the mean $p_T$ is increased marginally. For example, in a central collision with $N_{part}$=350, for all the species, mean $p_T$ is increased by $\sim$ 10\% from RHIC to LHC energy.   
    
\section{$p_T$-spectra  in Pb+Pb collisions at LHC}
 
\begin{figure}[t]
\includegraphics[bb=27 289 527 770
,width=0.5\linewidth,clip]{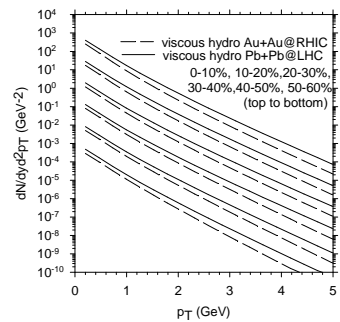}
\caption{The solid lines are minimally viscous hydrodynamic predictions
for $\pi^-$ $p_T$ spectra in 0-10\%, 10-20\%, 20-30\%, 30-40\%, 40-50\% and 50-60\% centrality Pb+Pb collisions at LHC. The dashed lines are the same for Au+Au collisions at RHIC.
To account for neglect of resonance production, the yield are normalised by a factor of 1.4.}
\label{F4}
\end{figure}

Minimally viscous hydrodynamic model predictions for $\pi^-$ $p_T$-spectra, in
0-10\%, 10-20\%, 20-30\%, 30-40\%, 40-50\% and 50-60\%  centrality   Pb+Pb collisions at LHC are shown in Fig.\ref{F4}.  We have neglected resonance
production.  
Resonance decay contribute to particle yield, more at low $p_T$ than at large $p_T$.    For example, at freeze-out temperature $T_F$=150 MeV, nearly $\sim$ 50\% of total pions are from resonance decay at $M_T$=0.5 GeV, contribution of decay pions decreases to $\sim$20\% at higher $M_T$=2 GeV \cite{Heinz:2004qz}. 
Minimally viscous hydrodynamics predictions, normalised by an factor of N=1.4, well explained the $\pi^-$ $p_T$ spectra in Au+Au collisions.   However, we must mention that overall normalisation do not account correctly for the resonance contribution to particle $p_T$-spectra. $p_T$ spectra will be uncertain by 10-15\%.  For comparison, in Fig.\ref{F4}, predicted spectra in Au+Au collisions at RHIC are shown as the dashed lines.
$p_T$ spectra are slightly flattened at LHC.   It is consistent with the predicted small increase in mean $p_T$ (see Fig.\ref{F3}) in LHC energy.

\begin{figure}[t]
\includegraphics[bb=26 290 527 770
,width=0.5\linewidth,clip]{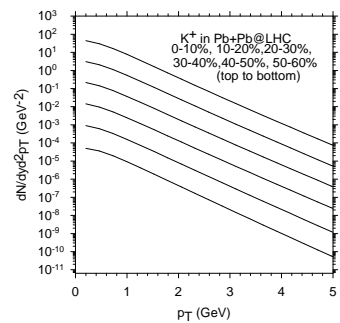}
\caption{ The   solid lines are minimally viscous hydrodynamics predictions for the   $K^+$ invariant yield in Pb+Pb collisions at LHC energy.}
\label{F5}
\end{figure}

\begin{figure}[h]
\includegraphics[bb=26 290 527 770
,width=0.5\linewidth,clip]{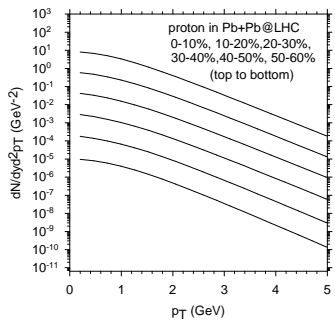}
\caption{same as in Fig.\ref{F5} but for protons.}
\label{F6}
\end{figure}

In Fig.\ref{F5} and \ref{F6}, we have shown the predicted $p_T$ spectra for $K^+$ and protons. We have not shown, but here again, compared to RHIC energy $p_T$ spectra is flattened.  
Before we digress, we would like to note that even though we
 have shown predictions right up to $p_T$=5 GeV, at large $p_T >$ 3 GeV, there may be other sources (e.g. jets) for particle production.  It is unlikely that viscous dynamics will predict correctly particle production at high $p_T >$ 3GeV. Present predictions are expected to be in reasonable agreement with future experiments up to $p_T \sim$ 3 GeV.  

\section{Elliptic flow in Pb+Pb collisions at LHC}

One of the important observations in Au+Au collisions at RHIC is the significant elliptic flow in non-central collisions. Qualitatively, elliptic flow is   explained in a hydrodynamic model, re-scattering of secondaries generates pressure and drives the subsequent collective motion. In non-central collisions, the reaction zone is asymmetric (almond shaped), pressure gradient is large in one direction and small in the other. The asymmetric pressure gradients generate the elliptic flow. As the fluid evolve and expands, asymmetry in the reaction zone decreases and  comes a stage when the reaction zone become symmetric and system no longer generate elliptic flow.  Elliptic flow is an early time phenomena and a  sensitive probe to the early stage of the fluid.

\begin{figure}[t]
\includegraphics[bb=46 291 524 769
,width=0.5\linewidth,clip]{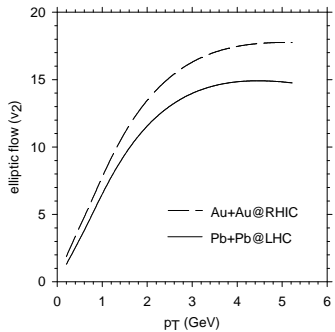}
\caption{ 
The solid line is the minimally viscous hydrodynamics prediction for
the minimum bias elliptic flow in Pb+Pb collisions at LHC. For comparison,
minimum bias elliptic flow in Au+Au collisions at RHIC is also shown (the dashed line). }
\label{F7}
\end{figure}

In Fig.\ref{F7}, solid line is the predicted elliptic flow in minimum bias Pb+Pb collisions at LHC. 
For comparison, we have shown the minimum bias elliptic flow in Au+Au collisions at RHIC (the dahsed line). It is not shown here, but 
experimental minimum bias $v_2$ in Au+Au collisions is well reproduced in
minimally viscous hydrodynamics.  
It is interesting to note that in Pb+Pb collisions, elliptic flow is reduced  by $\sim$ 15\%.  Reduction of elliptic flow in LHC has also been predicted by Krieg and Bleicher \cite{Krieg:2007sx}. In a parton recombination model, they studied the energy dependence of elliptic flow in heavy ion collisions from AGS to LHC energy. It was observed that from RHIC to LHC  energy elliptic flow decreases. Parton transport models also predict less elliptic flow in LHC energy than in Au+Au collisions at RHIC \cite{Molnar:2007an}.

\begin{figure}[t]
\includegraphics[bb= 17 216 567 769
,width=0.5\linewidth,clip]{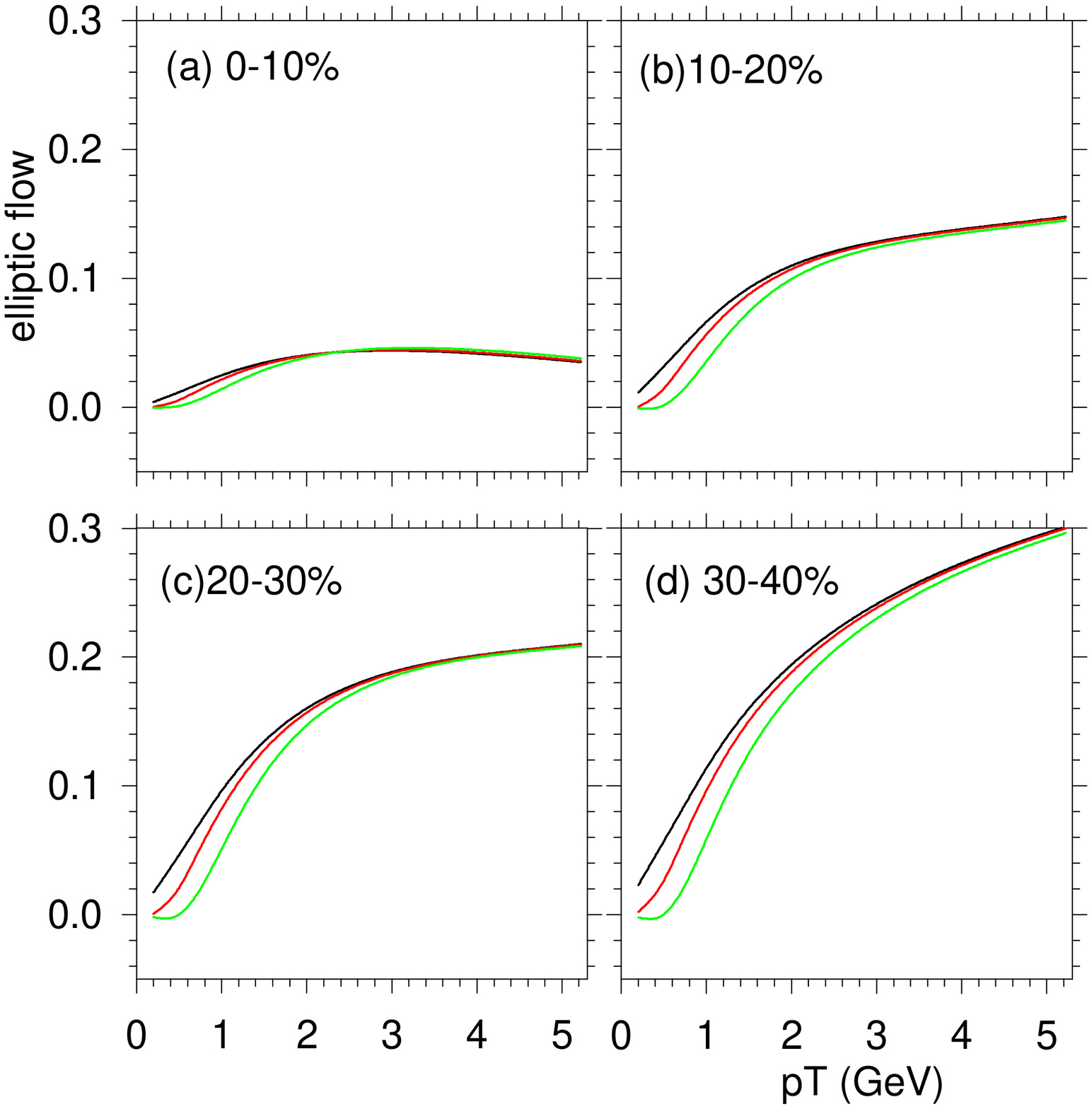} 
\caption{(color online) Predicted elliptic flow in 0-10\%, 10-20\%, 20-30\% and 30-40\% Pb+Pb collisions at LHC are shown. The black, red and green lines are  elliptic flow   for $\pi^-$, $K^+$ and protons respectively.} 
\label{F8}
\end{figure}

In Fig.\ref{F8}, in four panels, we have shown the minimally viscous dynamics predictions for elliptic flow in   0-10\%, 10-20\%, 20-30\% and, 30-40\%   centrality Pb+Pb collisions at LHC. The black, red and green lines corresponds to elliptic flow for $\pi^-$, $K^+$ and protons. Species dependence is similar to that at RHIC energy. At large $p_T$, elliptic flow between different species are marginally different. Species dependence is seen only at low $p_T$, lighter the particle, more is the elliptic flow.   
We have not shown any comparison with $v_2$ at RHIC. But as with the minimum bias collisions, elliptic flow  
in different centrality ranges of collisions is reduced at LHC. 

\section{Summary}

To summarise, in a minimally viscous ($\eta/s$=0.08) hydrodynamics , we have given predictions for several experimental observables in Pb+Pb collisions at LHC energy ($\sqrt{s}$=5.5 TeV). Assuming that particle multiplicity increase logarithmically with c.m. energy, we have extrapolated the lower energy  data (tabulated by the PHENIX collaboration) to obtain an estimate of particle multiplicity at LHC energy collisions. Our estimate indicate that a Pb+Pb collision with participant number  $N_{part}$=350, produces $\sim$ 927 charged particles.
The initial central entropy density of the fluid ($S_{ini}$=180 $fm^{-3}$) in Pb+Pb collisions was fixed to reproduce the extrapolated particle multiplicity.  
The initial time ($\tau_i$) and freeze-out temperature ($T_F$)
 were kept fixed at the value, $\tau_i$=0.6 fm and $T_F$=130 MeV, as it was obtained in the analysis of Au+Au data in RHIC energy collisions.

Our predictions indicate that compared to Au+Au collisions at RHIC, in Pb+Pb collisions at LHC, (i) particle multiplicity will increase by a factor of $\sim$ 1.6, (ii) the mean $p_T$ will be enhanced by $\sim$ 10\%, (iii) $p_T$ spectra of identified particles will be (slightly) flattened and (iv) 
elliptic flow will decrease  by $\sim$ 15\%.  We hope the results will be helpful in planning future experiments in LHC energy.

\end{document}